\documentclass[conference]{IEEEtran}
\IEEEoverridecommandlockouts

\usepackage{cite}
\usepackage{amsmath,amssymb,amsfonts}
\usepackage{algorithmic}
\usepackage{graphicx}
\usepackage{textcomp}
\usepackage{xcolor}
\usepackage{booktabs}
\usepackage{marvosym}
\def\BibTeX{{\rm B\kern-.05em{\sc i\kern-.025em b}\kern-.08em
    T\kern-.1667em\lower.7ex\hbox{E}\kern-.125emX}}
\begin{document}

\title{Beyond H\&E: Unlocking Pathological Insights with Polarization Imaging\\
}

\DeclareRobustCommand*{\IEEEauthorrefmark}[1]{%
  \raisebox{0pt}[0pt][0pt]{\textsuperscript{\footnotesize #1}}%
}

\author{
    \IEEEauthorblockN{
        Yao Du\IEEEauthorrefmark{1}, 
        Jiaxin Zhuang\IEEEauthorrefmark{1},
        Xiaoyu Zheng\IEEEauthorrefmark{1},
        Jing Cong\IEEEauthorrefmark{2}, 
        Limei Guo\IEEEauthorrefmark{3}, 
        Chao He\IEEEauthorrefmark{4 \Letter}, 
        Lin Luo\IEEEauthorrefmark{5 \Letter}, and 
        Xiaomeng Li\IEEEauthorrefmark{1 \Letter}
    }
    \thanks{\textsuperscript{\Letter} Corresponding author: Chao He, Email: chao.he@eng.ox.ac.uk;
    Lin Luo, Email: luol@pku.edu.cn; Xiaomeng Li, Email: eexmli@ust.hk}

    \IEEEauthorblockA{
        \IEEEauthorrefmark{1} The Hong Kong University of Science and Technology, Hong Kong SAR, China\\
        \IEEEauthorrefmark{2} Beijing Institute of Collaborative Innovation, China \\
        \IEEEauthorrefmark{3} Peking University Health Science Center, Peking University Third Hospital, China \\
        \IEEEauthorrefmark{4} University of Oxford, United Kingdom \\
        \IEEEauthorrefmark{5} Peking University, China \\
    }
}


\maketitle

\begin{abstract}
Histopathology image analysis is fundamental to digital pathology, with hematoxylin and eosin (H\&E) staining as the gold standard for diagnostic and prognostic assessments.
While H\&E imaging effectively highlights cellular and tissue structures, it lacks sensitivity to birefringence and tissue anisotropy, which are crucial for assessing collagen organization, fiber alignment, and microstructural alterations—key indicators of tumor progression, fibrosis, and other pathological conditions.
To bridge this gap, we construct a polarization imaging system and curate a new dataset of over 13,000 paired Polar-H\&E images. Visualizations of polarization properties reveal distinctive optical signatures in pathological tissues, underscoring its diagnostic value.
Building on this dataset, we propose \textbf{PolarHE}, a dual-modality fusion framework that integrates H\&E with polarization imaging, leveraging the latter’s ability to enhance tissue characterization.
Our approach employs a feature decomposition strategy to disentangle common and modality-specific features, ensuring effective multimodal representation learning.
Through comprehensive validation, our approach significantly outperforms previous methods, achieving an accuracy of 86.70\% on the Chaoyang dataset and 89.06\% on the MHIST dataset. 
These results demonstrate that polarization imaging is a powerful and underutilized modality in computational pathology, enriching feature representation and improving diagnostic accuracy. \textbf{PolarHE} establishes a promising direction for multimodal learning, paving the way for more interpretable and generalizable pathology models.
\end{abstract}


\section{Introduction}
\label{sec:intro}

Histopathology image analysis is fundamental in digital pathology, with hematoxylin and eosin (H\&E) staining serving as the gold standard for diagnosing and prognosticating various diseases, particularly cancer~\cite{xiang2025moc}. 
While H\&E imaging highlights cellular and tissue structures, it lacks the ability to capture birefringence and tissue anisotropy, which provide critical insights into collagen organization, fiber alignment, and microstructural alterations—key indicators of tumor progression, fibrosis, and other pathological conditions~\cite{dong2021polarization,zhang2024incidence,zheng2025diffusion}.


Polarization imaging, which leverages the fundamental properties of light, has emerged as a promising auxiliary modality for histopathological analysis.
By capturing tissue birefringence, scattering properties, and depolarization effects, it provides additional contrast that reveals microstructural and biochemical changes beyond what is visible in standard H\&E staining~\cite{he2021polarisation}. 
More specifically, polarization imaging enables non-invasive acquisition of birefringence characteristics and sub-resolution structural information by analyzing how biological tissues modulate the polarization state of incident light waves—typically quantified through Stokes vectors, Mueller matrices, and derived optical parameters~\cite{hao2024cartesian,he2021polarisation}. This imaging mechanism demonstrates high sensitivity to features such as collagen fiber alignment and cell membrane integrity, which are critical indicators of pathological processes.

Recent studies have validated the clinical potential of polarization imaging across diverse applications, including tumor boundary delineation, fibrosis evaluation, cervical intraepithelial neoplasia grading, and breast carcinoma classification~\cite{wang2024pathology, pati2024Accelerating, dong2021Deriving, dong2021polarization, hou2022polarimetry}. By enabling high-accuracy distinction of pathological features such as collagen alignment and nuclear organization, polarization imaging emerges as a powerful complement to traditional histopathology—particularly for early disease detection and precise tissue characterization.

Despite its potential, polarization imaging remains underutilized in computational pathology, primarily due to the lack of established datasets and methodologies for its integration with conventional H\&E-stained slides. 
To bridge this gap, we construct a polarization imaging system and collect a novel dataset of paired H\&E-stained and polarization images. It comprises over 13,000 spatially aligned image patches, capturing both conventional histological structure and polarization-based birefringence features. This dataset provides a rich foundation for exploring complementary modality representations in tissue analysis.
Polarization property visualization confirms polarization captures distinct optical tissue signatures, reinforcing the complementary role of polarization imaging in pathology.

As discussed above, H\&E and polarization images offer complementary information for clinical pathology. Effectively leveraging this complementarity is crucial, especially considering that polarization imaging is costly and less accessible in routine clinical practice. To address this, we propose a dual-modality fusion strategy that integrates both H\&E and polarization data during training. This enables the model to learn richer, more informative representations, allowing the resulting encoders to generalize better on downstream tasks.
Recent advancements in self-supervised learning (SSL) offer a promising avenue for multimodal fusion without requiring extensive manual annotations~\cite{nguyen2024towards}.
SSL frameworks, which learn meaningful representations from unlabeled data by solving pretext tasks, have shown remarkable success across various domains, including medical imaging~\cite{kang2023benchmarking,koohbanani2021self,wu2024voco,dai2023semi}.
By jointly learning from H\&E and polarization imaging in a self-supervised manner, it becomes possible to develop robust representations that capture both shared and modality-specific features, enhancing downstream tasks such as patch classification and tissue segmentation~\cite{wang2024decoupling}.

Therefore, we propose \textbf{PolarHE}, a dual-modality fusion framework that integrates H\&E and polarization imaging for enhanced representation learning in pathology.
Our framework employs a feature decomposition strategy to disentangle common and modality-specific features, ensuring shared features remain aligned across modalities while preserving the discriminative properties of unique modality features.
We validate our framework on a patch-level classification task, where PolarHE significantly outperforms previous methods.

Our key contributions are as follows: 1) We construct the first dual-modality Polarization Pathology Dataset, which includes over 13,000 image patches with precise spatial registration between H\&E and polarization imaging;
2) We introduce \textbf{PolarHE}, a novel dual-modality fusion framework combining H\&E and polarization imaging for enhanced histopathology analysis. We adopt a feature decomposition strategy that disentangles common and modality-specific features, facilitating multimodal learning; 
3) We demonstrate model effectiveness through extensive experiments, showing significant performance improvements on patch classification tasks.

\section{Polarization System and Paired Dataset Construction}

\begin{figure}[t]
\begin{center}
 \scalebox{0.48}{
   \includegraphics[width=\textwidth]{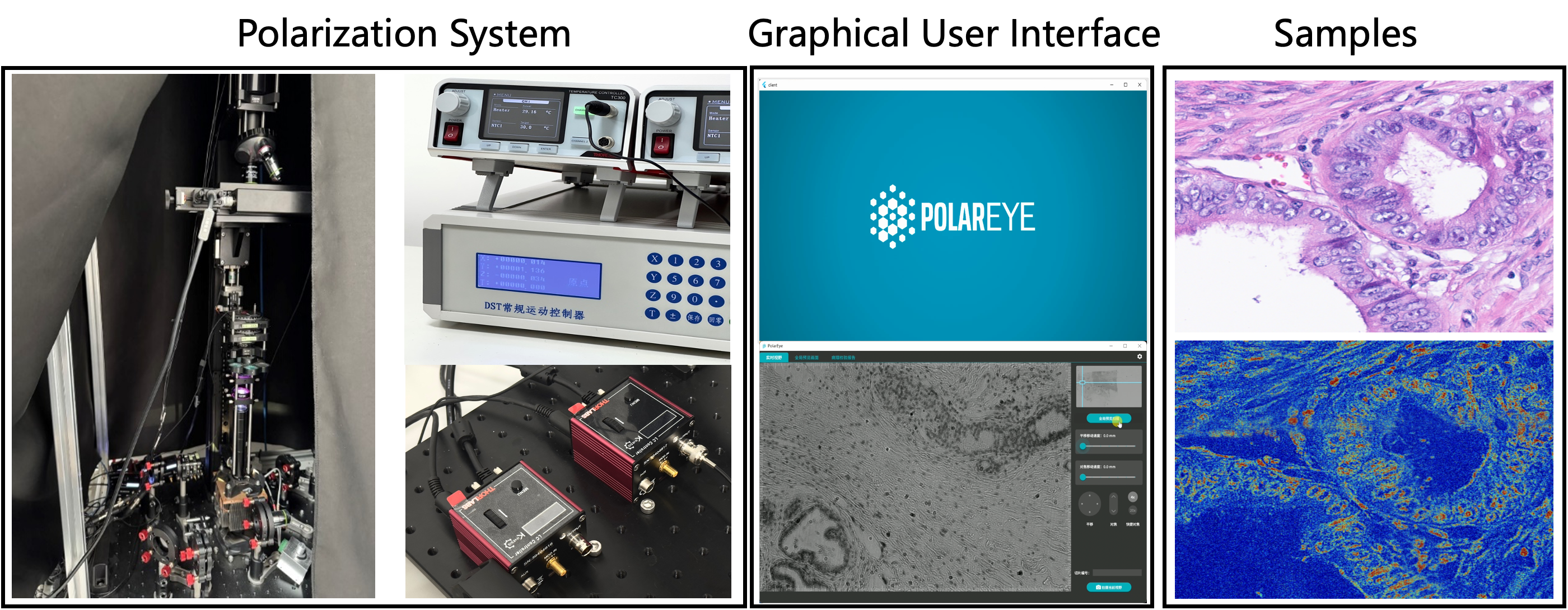}
   }
\end{center}
\vspace{-5mm}
   \caption{Polarization Imaging Acquisition and Processing System.
   }
\vspace{-7mm}
\label{fig:system}
\end{figure}

\noindent\textbf{Polarization Imaging System for Pathology.}
To facilitate multimodal pathology image analysis, we construct a polarization imaging system for data acquisition and processing, as illustrated in Figure \ref{fig:system}. 
This system is designed to capture full-vector polarization imaging, enabling comprehensive tissue characterization beyond H\&E staining.

The acquisition process begins with the use of specialized slide scanners to obtain Whole-Slide Images (WSIs) of H\&E staining. Subsequently, the samples are placed on a stage for polarimetric imaging.
We employ a 635nm red light source as the incident light and perform full-field scanning at 4× magnification to ensure uniform tissue coverage across all pathology slides. 
After autofocus determines the optimal focal plane, the system scans the sample region-by-region with a preset step size. Each field of view undergoes polarization measurement, generating a 16-dimensional polarization matrix $M_{s(u,v)}$ for each pixel $(u,v)$.
This results in a 16-channel polarization representation, providing rich structural and birefringence information, which is essential for analyzing tissue anisotropy and microstructural organization.

Given that samples are large and a single
shot cannot cover the entire slice, an ITKMontage~\cite{zukic2021itkmontage} module is employed for multi-modality image registration and stitching. To address issues such as size consistency and missing data in multi-modality image stitching, the system automatically generates configuration files to record tile position information. It also utilizes boundary detection and virtual tile generation techniques. Additionally, grayscale uniformity correction and brightness normalization are applied. Ultimately, this process results in the formation of complete panoramic WSI images under a 4× objective lens.

\noindent\textbf{Spatial Alignment via Image Registration.}
To establish a precisely aligned multimodal dataset, we store the corresponding H\&E-stained WSIs alongside the polarization images, preserving their structural and morphological details. To enable pixel-wise comparison across modalities, we design a multi-stage registration pipeline that ensures precise spatial alignment.
Initial pre-processing standardizes the image scales and removes extraneous blank regions to mitigate feature interference.
The polarization-derived grayscale image is used as the registration reference due to its structural consistency with other modalities. 
Feature matching is performed using the SuperPoint algorithm~\cite{detone2018superpoint}, a deep learning-based detector that extracts cross-modal keypoints for robust correspondence estimation. 
Rigid registration aligns global structures via affine transformations. Subsequent non-rigid registration refines local deformations using B-spline
modeling~\cite{unser2002b} combined with curvature regularization~\cite{myronenko2006non} to prevent artifacts.
After registration, all modalities are resampled to 2304×1296 pixels and aligned
with the polarization field-of-view (FOV). Adaptive thresholding is then applied to exclude tissue-free or artifact-dominated regions. 
The validated FOVs are partitioned into 224×224 pixel patches using a sliding window approach, preserving multi-modality alignment through strict spatial correspondence.

This process yields a multimodal dataset comprising over 13,000 high-quality, spatially aligned patches.
This dataset enables integrated learning from H\&E and polarization modalities, providing a foundation for investigating multimodal fusion strategies and advancing downstream pathology analysis.

\noindent\textbf{Polarization Property Visualization.}
Polarization imaging offers a powerful contrast mechanism for visualizing tissue microstructure and biochemical composition, enabling fine-grained differentiation of pathological features. This imaging modality derives pixel-wise optical parameters through Mueller matrix decomposition, capturing subtle light–tissue interactions. The key extracted parameters include:
\begin{figure}[t]
\begin{center}
\scalebox{0.48}{
   \includegraphics[width=\textwidth]{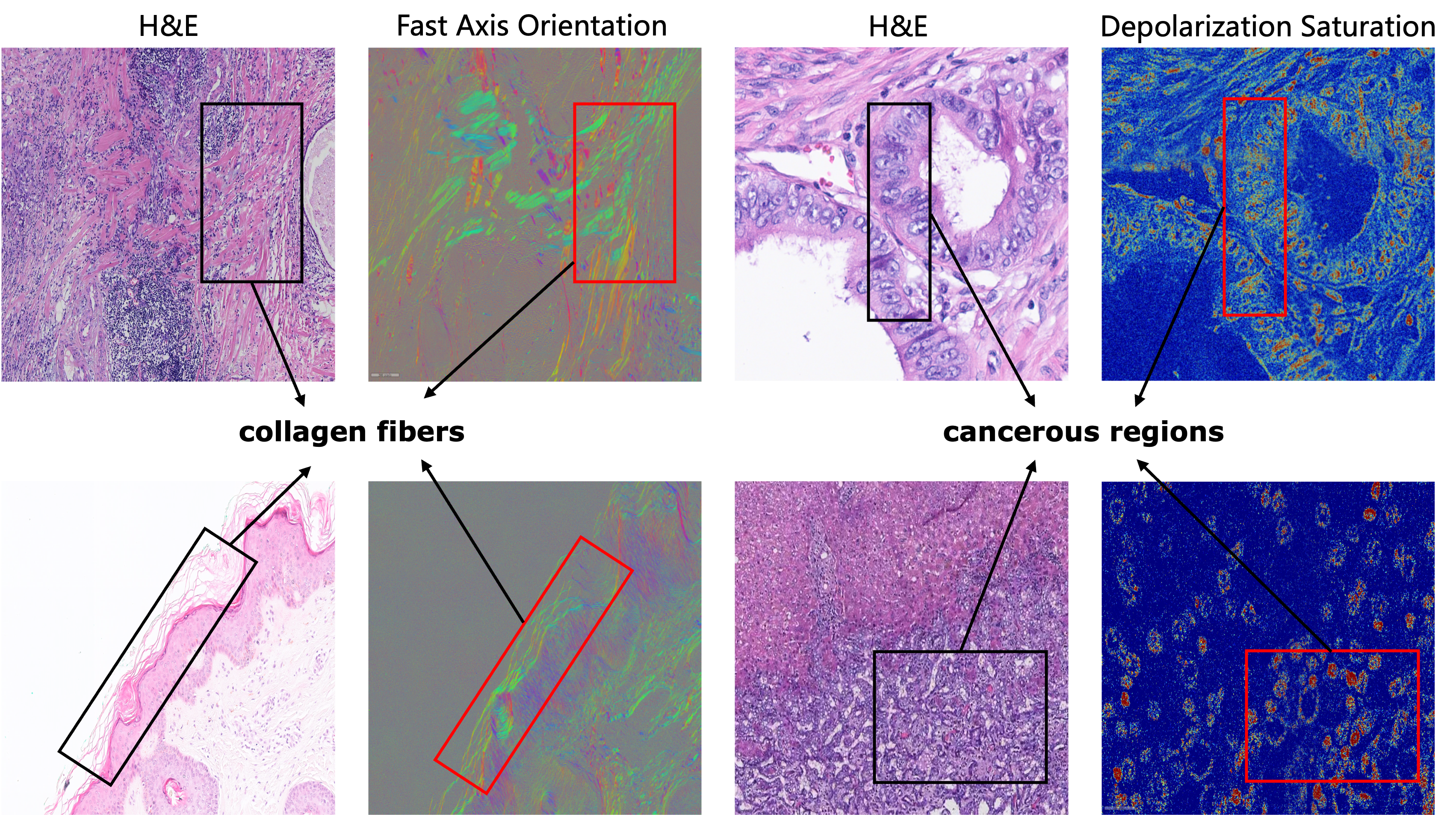}
   }
\end{center}
\vspace{-5mm}
   \caption{Polarization Property (Fast-axis Orientation and Depolarization Saturation)
   }
\vspace{-5mm}
\label{fig:polar matrix}
\end{figure}
1) Phase Retardation: quantifies the phase shift between orthogonal polarization components introduced by birefringent structures such as collagen fibers. It serves as an indicator of microstructural anisotropy and fiber density;
2) Fast-axis Orientation: represents the principal axis of birefringence at each pixel, revealing the alignment and directional organization of fibrous tissue components;
3) Depolarization Power: measures the extent to which polarized light becomes randomized after passing through tissue. Higher depolarization is typically associated with structural heterogeneity, such as densely packed or disorganized cell nuclei~\cite{yang2024data}.

\begin{figure*}[ht]
\begin{center}
   \scalebox{0.70}{
   \includegraphics[width=\textwidth]{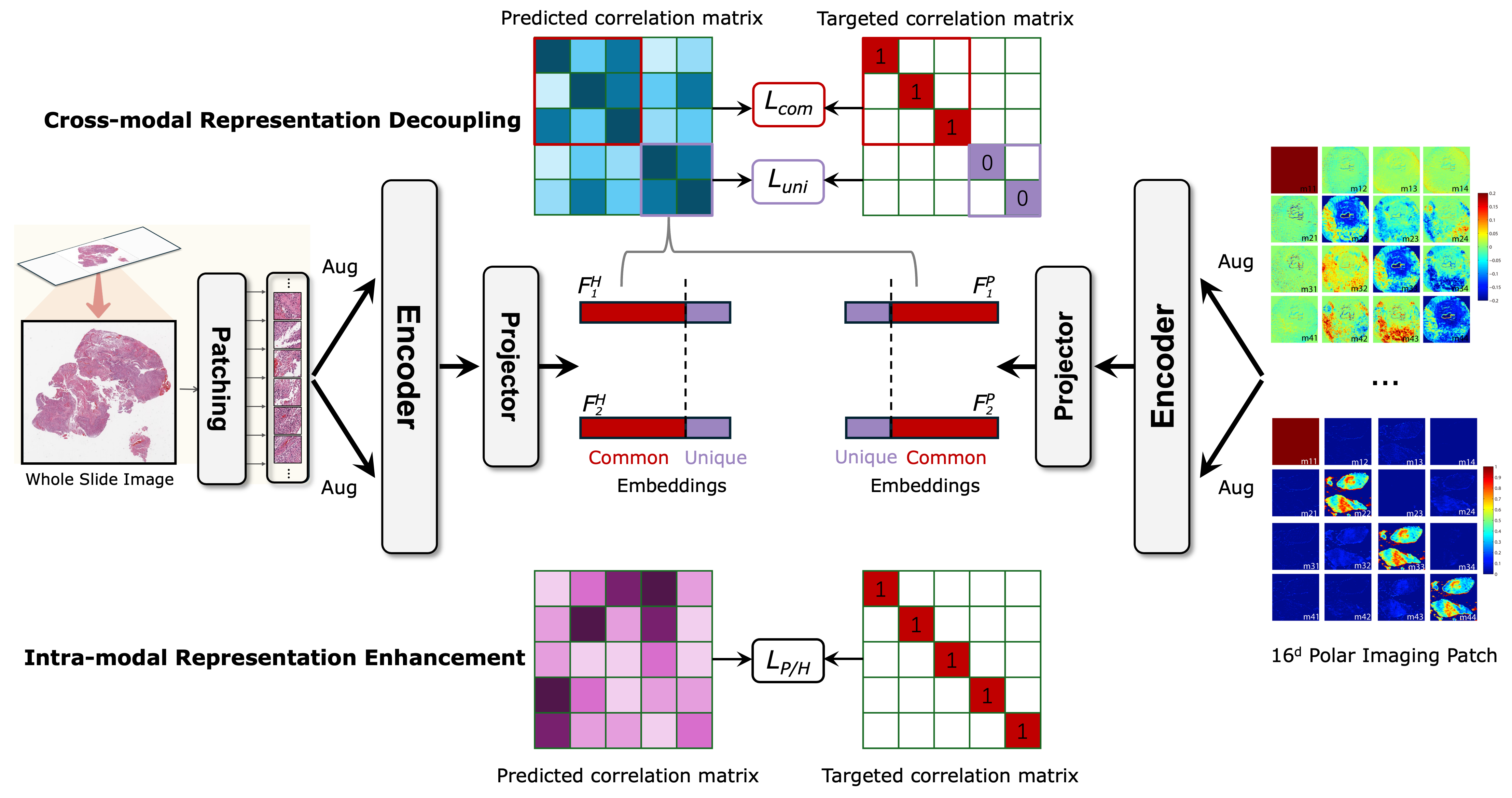}
   }
\end{center}
\vspace{-5mm}
   \caption{Integrating H\&E and polarization representations via feature decomposition.
   }
\vspace{-6mm}
\label{fig:decur}
\end{figure*}

As shown in the left part of Figure~\ref{fig:polar matrix}, compared to traditional H\&E staining—where collagen fibers often appear pink and are visually indistinguishable from muscle or blood vessels—polarization imaging, particularly through fast-axis orientation maps, enables clear delineation of collagen structures. This allows for more accurate assessment of fiber arrangement and tissue integrity, which is especially valuable in detecting fibrosis or tumor infiltration.

In the right part of Figure~\ref{fig:polar matrix}, pathological cases such as endometrioid carcinoma exhibit distinctive polarization signatures. Compared to normal hepatocytes, cancerous tissues show elevated DNA replication, increased nuclear stiffness, and stronger depolarization signals, likely reflecting chromatin condensation and cellular heterogeneity. Additionally, cilia located along cavity edges may display polarization-sensitive characteristics. These features indicate variations in depolarization saturation can serve as optical biomarkers for identifying malignant regions and evaluating tumor progression.

Together, these observations underscore the unique capability of polarization imaging to extract interpretable and quantitative optical markers—providing a powerful complement to conventional H\&E analysis and substantially enhancing diagnostic resolution in digital pathology.


\section{Methodology}
\label{sec: method}

\noindent\textbf{Overview.}
Unlike existing multimodal SSL methods that typically operate on loosely or semantically aligned modalities~\cite{wang2024decoupling, zbontar2021barlow}, our framework is tailored for the histopathology domain, where precise spatial alignment between H\&E and polarization images enables meaningful feature-level fusion. While the input modalities are pixel-aligned, we deliberately perform fusion at the representation level to ensure flexibility and generalizability.
By exploiting their complementary characteristics—H\&E capturing morphology and polarization revealing sub-resolution optical cues—we design a polarization-guided training strategy that enriches H\&E representation learning. This allows the H\&E encoder to absorb polarization-driven structural cues, improving performance even when only H\&E is available at inference. To the best of our knowledge, this is the first SSL framework to jointly learn from H\&E and polarization imaging in digital pathology.

To learn joint representations from H\&E and polarization images, we introduce \textbf{PolarHE},
a modality decomposition framework that explicitly separates shared and modality-specific features in a self-supervised manner. This enables the model to preserve complementary information from both imaging modalities, improving robustness and generalization.
As illustrated in Figure \ref{fig:decur}, our framework processes two augmented views of each modality through modality-specific encoders and projectors, generating the respective embeddings: ${F^H_{1}}, {F^H_{2}}$ for H\&E and ${F^P_{1}}, {F^P_{2}}$ for polarization images.
Batch normalization is applied to align feature distributions, ensuring consistency across the dataset.
To separate common and unique modality-specific representations, we leverage cross-correlation matrices for feature disentanglement, enforcing alignment for shared features while preserving discriminative properties in modality-unique components.


\noindent\textbf{Cross-correlation Matrix for Feature Dependency Analysis.} 
To measure dependencies between embedding dimensions, we compute cross-correlation matrices.
Given two embedding vectors  $F^{A}, F^{B} \in \mathbb{R}^{K}$, correlation matrix $\mathcal{C}$  is defined as:

\begin{small}
\begin{equation}
\mathcal{C}_{i j} = \frac{\sum_b f_{b, i}^A f_{b, j}^B}{\sqrt{\sum_b\left(f_{b, i}^A\right)^2} \sqrt{\sum_b\left(f_{b, j}^B\right)^2}}\quad
\end{equation}
\end{small}

\noindent where $b$ indexes batch samples, and $i$, $j$ index embedding dimensions.
The resulting square matrix $\mathcal{C} \in \mathbb{R}^{K \times K}$ contains values ranging from -1 to 1, quantifying feature correlation across different dimensions.
In the cross-modal case, the correlation matrix $\mathcal{C}$ is calculated between two embeddings from different modalities, such as ${F^H_{1}}$ and ${F^P_{1}}$ in Figure~\ref{fig:decur}.

\noindent\textbf{Cross-modal Representation Decoupling.} 
To ensure proper feature disentanglement, we partition the embedding space $K$ into: $K_c$ (common representations) and $K_u$ (unique representations) with $K_c+K_u=K$. 
In Figure \ref{fig:decur}, the common representations are identical (highlighted in red) and encode modality-invariant features, while the modality-unique representations are decorrelated (highlighted in purple) and capture specialized modality-dependent structures.

To enforce cross-modal common feature alignment, we extract a sub-matrix $\mathcal{C_\text{c}} \in \mathbb{R}^{K_c \times K_c}$ from the common dimensions of ${F^H_{1}}$ and ${F^P_{1}}$. The cross-modal redundancy reduction loss is formulated as:

\begin{small}
\begin{equation}
\mathcal{L}_{com} = \sum_i\left(1-\mathcal{C_\text{c}}_{i i}\right)^2+\lambda_{c} \cdot \sum_i \sum_{j \neq i} \mathcal{C_\text{c}}_{i j}^2\quad,
\end{equation}
\end{small}
\noindent where $\lambda_{c}$ is a weighting coefficient balancing invariance enforcement (to ensure common features remain aligned across modalities) and decorrelation (to reduce feature redundancy and avoid model collapse).

For modality-unique representations, we extract a separate sub-matrix $\mathcal{C_\text{u}} \in \mathbb{R}^{K_u \times K_u}$ from ${F^H_{1}}$ and ${F^P_{1}}$. The objective here is to decorrelate the unique dimensions, encouraging them to retain distinct modality-specific information:

\begin{small}
\begin{equation}
\mathcal{L}_{uni} = \sum_i\mathcal{C_\text{u}}_{i i}^2+\lambda_{u} \cdot \sum_i \sum_{j \neq i} \mathcal{C_\text{u}}_{i j}^2 \quad,
\end{equation}
\end{small}
\noindent where $\lambda_{u}$ regulates the balance between inter-modal decorrelation and redundancy reduction. 
However, simply enforcing decorrelation does not guarantee that the unique embeddings retain meaningful pathology-related features, as they may collapse into random, decorrelated noise. To address this issue, we introduce intra-modal representation enhancement.

\noindent\textbf{Intra-modal Representation Enhancement.} 
To reinforce feature stability and prevent collapse of decoupled unique dimensions, we introduce intra-modal training that covers all the embedding dimensions, ensuring that the model learns robust within-modality representations.
For each modality, a cross-correlation matrix $\mathcal{C_\text{H}}$ (or $\mathcal{C_\text{P}}$) is generated from the full dimensions of the embedding vectors ${F^H_{1}}$ and ${F^H_{2}}$ (or ${F^P_{1}}$ and ${F^P_{2}}$), optimizing: 
\begin{small}
\begin{equation}
\mathcal{L}_{H} = \sum_i\left(1-\mathcal{C_\text{H}}_{i i}\right)^2+\lambda_{H} \cdot \sum_i \sum_{j \neq i} \mathcal{C_\text{H}}_{i j}^2 \quad,
\end{equation}
\end{small}

\noindent where $\lambda_{H}$ regulate the balance between feature invariance and redundancy minimization. This prevents degeneration and ensures intra-modal consistency. Similar implementation for ${F^P_{1}}$ and ${F^P_{2}}$ to obtain $\mathcal{L}_{P}$.

\noindent \textbf{Overall training objective}.
The final self-supervised learning objective integrates cross-modal common representation learning, modality-unique decomposition, and intra-modal consistency constraints:
\begin{small}
\begin{equation}
\mathcal{L} = \mathcal{L}_{com} + \mathcal{L}_{uni} + \mathcal{L}_{H} + \mathcal{L}_{P}\quad.
\end{equation}
\end{small}
\noindent This ensures modality-invariant feature alignment, modality-specific feature decorrelation, and intra-modal representation reinforcement.
By incorporating these objectives, \textbf{PolarHE} effectively fuses H\&E and polarization imaging, leading to richer pathology representations and superior downstream performance.

\section{Experiments}

\noindent \textbf{Implementation.}
Each branch of PolarHE consists of a dedicated backbone and a three-layer MLP projector, each with an output dimension of 8192. We adopt the UNI ViT-L/16 encoder~\cite{chen2024towards} as the backbone,  as it provides a strong H\&E-specific representation baseline, which not only facilitates effective integration with polarization features but also ensures that observed gains can be attributed to our proposed fusion strategy rather than backbone selection.
Training is performed on embedding representations obtained after the projector, where the feature space is partitioned into common and unique components, with 75\% allocated to common features. During training, both encoder and projector are jointly optimized with a small learning rate of 1e-5. We apply data augmentations including random horizontal/vertical flipping and rotation. The encoder is subsequently transferred to downstream tasks, while the trade-off parameter ($\lambda$) for the loss terms is set to 0.0051.


\noindent\textbf{Downstream Patch Classification.}
To assess the effectiveness of PolarHE, we conduct patch classification experiments on two benchmark histopathology datasets:

\noindent 1. Chaoyang Dataset~\cite{zhu2021hard} consists of colon tissue slides collected from Chaoyang Hospital.
Each patch is 512 × 512 pixels, with the dataset split as follows: Training Set: 1,111 normal, 842 serrated, 1,404 adenocarcinoma, and 664 adenoma patches; Testing Set: 705 normal, 321 serrated, 840 adenocarcinoma, and 273 adenoma patches. 

\noindent 2. MHIST Dataset~\cite{wei2021petri} consists of 2,175 training images and 977 testing images from 328 WSIs. These images are labeled into two classes: Hyperplastic Polyps (HPs) and Sessile Serrated Adenomas (SSAs). 
From these WSIs, 3,152 diagnostically relevant patches of size 224 × 224 pixels are extracted,representing clinically significant tissue regions.

\begin{table}[ht]
    \centering
    \renewcommand{\arraystretch}{1.2}
    \setlength{\tabcolsep}{6pt}
    \vspace{-3mm}
    \caption{Test performance on Chaoyang~\cite{zhu2021hard} dataset.}
    \vspace{-3mm}
    \scalebox{0.80}{
    \begin{tabular}{lccc}
        \toprule
        \textbf{Method} & \textbf{Acc} & \textbf{F1} & \textbf{AUC} \\
        \midrule
        DINO~\cite{caron2021emerging} & 80.21 & - & - \\
        HSA-NRL~\cite{zhu2021hard} & 83.40 & 76.54 & 94.51 \\
        SSAT~\cite{das2024limited} & 82.52 & - & - \\
        ViT-DAE~\cite{xu2023vit} & 80.12 & 78.60 & - \\
        INCV~\cite{chen2019understanding} & 80.34 & 74.11 & 92.63 \\
        2HSL~\cite{galdran2023multi} & 82.90 & - & - \\
        OUSM~\cite{xue2019robust} & 80.53 & 73.70 & 93.69 \\
        SELF~\cite{nguyen2019self} & 80.49 & 75.31 & 93.99 \\
        DiRL~\cite{kapse2024attention} & 83.60 & - & 95.01 \\
        SH-PEFT~\cite{liu2024sparsity} & 84.80 & 80.61 & - \\
        UNI~\cite{chen2024towards} & 85.51 & 85.29 & 95.21 \\
        \midrule
        \textbf{PolarHE (ours)} & \textbf{86.70} & \textbf{86.42} & \textbf{95.96} \\
        \bottomrule
    \end{tabular}
    }
    \vspace{-6mm}
    \label{tab:chaoyang}
\end{table}

\noindent\textbf{Patch Classification by Linear Probing.}
To evaluate the quality of the representations learned by PolarHE, we adopt a linear probing protocol, where a lightweight classifier is trained on top of the frozen encoder. This setup, following standard benchmarks~\cite{kang2023benchmarking}, isolates the representation quality from task-specific fine-tuning.

We report the average performance over three random seeds on both the Chaoyang and MHIST datasets to ensure result robustness. For fair benchmarking, we compare against results reported in prior work or reproduced using publicly available code and official configurations.

Our method, PolarHE, achieves state-of-the-art performance on both datasets. On the Chaoyang Dataset from Table~\ref{tab:chaoyang}, it achieves an accuracy of 86.70\%, an F1 score of 86.42\%, and an AUC of 95.96\%. On the MHIST Dataset from Table~\ref{tab:mhist}, it achieves an accuracy of 89.06\%, an F1 score of 88.97\%, and an AUC of 94.10\%. 

\begin{table}[ht]
    \centering
    \renewcommand{\arraystretch}{1.2}
    \setlength{\tabcolsep}{6pt}
    \vspace{-3mm}
    \caption{Test performance on MHIST~\cite{wei2021petri} dataset.}
    \vspace{-3mm}
    \scalebox{0.80}{
    \begin{tabular}{lccc}
        \toprule
        \textbf{Method} & \textbf{Acc} & \textbf{F1} & \textbf{AUC} \\
        \midrule
        DINO~\cite{caron2021emerging} & 83.10 & 81.60 & 89.78 \\
        BarlowTwins~\cite{zbontar2021barlow} & 81.27 & - & - \\
        MoCo-v2~\cite{kang2023benchmarking} & 85.88 & - & - \\
        Bitfit~\cite{zaken2021bitfit} & 82.40 & 80.86 & 89.60 \\
        SSF~\cite{lian2022scaling} & 82.29 & 80.85 & 89.29 \\
        RMSGD~\cite{hosseini2022exploiting} & 82.58 & - & - \\
        TPP~\cite{lei2024pre} & 83.42 & 82.21 & 91.38 \\
        DiRL~\cite{kapse2024attention} & 78.20 & - & 87.11 \\
        Virchow~\cite{vorontsov2023virchow} & 83.40 & 83.51 & - \\
        CTransPath~\cite{wang2021transpath} & 83.75 & - & - \\
        UNI~\cite{chen2024towards} & 85.63 & 85.57 & 91.98 \\
        \midrule
        \textbf{PolarHE (ours)} & \textbf{89.06} & \textbf{88.97} & \textbf{94.10} \\
        \bottomrule
    \end{tabular}
    }
    \vspace{-3mm}
    \label{tab:mhist}
\end{table}

Importantly, only H\&E images are used during both linear probing and inference. This demonstrates that our polarization-guided pretraining substantially enhances the H\&E encoder’s downstream performance—even when polarization data is unavailable at test time. This reflects real-world deployment settings, where polarization imaging is costly or limited in availability, yet its benefits can still be leveraged through training-time supervision.

\noindent \textbf{Ablation Study.}
Table~\ref{tab:loss-ablation} shows removing intra-modal training or modality decoupling reduces accuracy, confirming their importance. Without intra-modal training, performance drops, especially on Chaoyang, highlighting its role in reinforcing modality-specific features. Without modality decoupling, accuracy also declines, showing the need for feature separation. The full model performs best, validating both components. For common representation ratio in Table~\ref{tab:common-ratio-ablation}, the downstream performance increases and decreases smoothly along with the change of the percentage of common dimensions, which indicates the sparsity of the common embedding space.


\begin{table}[ht]
\centering
\vspace{-3mm}
\caption{Ablation study on loss components. We report the accuracy on Chaoyang and MHIST datasets.}
\vspace{-2mm}
\label{tab:loss-ablation}
\scalebox{0.8}{
\begin{tabular}{l|cc}
\toprule
\textbf{Loss Component} & \textbf{Chaoyang} & \textbf{MHIST} \\
\midrule
PolarHE (ours)            & \textbf{86.70}    & \textbf{89.06} \\
w/o intra                 & 84.15             & 86.25          \\
w/o decoup.              & 85.92             & 87.06          \\
w/o intra \& decoup.     & 85.51             & 85.63          \\
\bottomrule
\end{tabular}
}
\end{table}

\vspace{-7mm}

\begin{table}[ht]
\centering
\caption{Effect of common representation ratio on accuracy.}
\vspace{-2mm}
\label{tab:common-ratio-ablation}
\scalebox{0.8}{
\begin{tabular}{c|cc}
\toprule
\textbf{Common Ratio} & \textbf{Chaoyang} & \textbf{MHIST} \\
\midrule
50\%   & 84.95    & 87.70 \\
75\%   & \textbf{86.70} & \textbf{89.06} \\
85\%   & 86.27    & 88.33 \\
\bottomrule
\end{tabular}
}
\vspace{-3mm}
\end{table}

\section{Conclusion}

We propose \textbf{PolarHE}, a dual-modality fusion framework that integrates polarization imaging with H\&E staining for enhanced representation learning in pathology. 
By leveraging a feature decomposition strategy, our approach effectively disentangles common and modality-specific features, leading to improved multimodal representation learning. 
Through comprehensive experiments, we demonstrated that the integration of H\&E and polarization representations leads to significantly better performance compared to existing methods. 
These findings underscore the untapped potential of polarization imaging as a powerful auxiliary modality in computational pathology. By enhancing diagnostic accuracy and model robustness, \textbf{PolarHE} marks a promising step toward more interpretable and automated pathology analysis, paving the way for advancements in multimodal deep learning for clinical applications.

\section{Acknowledgment}
This work was partially supported by the Research Grants Council (RGC) of the Hong Kong Special Administrative Region, China (Project No. R6005-24); the Hong Kong Joint Research Scheme (JRS) of the National Natural Science Foundation of China (NSFC)/RGC (Project No. N\_HKUST654/24); and the NSFC (Grant No. 62306254).
We would like to thank the support of St John’s College, the University of Oxford, and the Royal Society University Research Fellowship (URF\textbackslash R1\textbackslash 241734) (C.H.). 


{\small
\bibliographystyle{IEEEtranS}
\bibliography{iclr2024_conference}
}

\end{document}